\documentclass[useAMS,usenatbib]{mn2e}
\usepackage{graphicx}
\usepackage{float}
\usepackage{placeins}

 \title[Galactic masers: kinematics, spiral structure and the disk dynamic state]
 {Galactic masers: kinematics, spiral structure and the disk dynamic state}
 \author[Milky Way masers]
 {
 A. S. Rastorguev$^{1,2,3}$ \thanks{E-mail: rastor@sai.msu.ru},
 N. D. Utkin$^{1,2,3}$,
 M. V. Zabolotskikh$^{1,2}$,
 A. K. Dambis$^{1,2}$,
 \newauthor A. T. Bajkova$^{4}$ and
 V. V. Bobylev$^{4}$\\ 
 $^{1}$Lomonosov Moscow State University, 1, Leninskie Gory, Moscow, 119991, Russia\\
 $^{2}$Sternberg Astronomical Institute, 13, Universitetskii prospect, Moscow, 119992, Russia\\
 $^{3}$Lomonosov Moscow State University, Faculty of Physics, 1, bld.2, Leninskie Gory, Moscow,  119992, Russia\\
 $^{4}$Central (Pulkovo) Astronomical Observatory, Russian Academy of Sciences,
       65/1 Pulkovskoye Chaussee, St.~Petersburg, 196140, Russia\\
 }

\begin{document}
 \date{Received 2016 March, 3; Accepted 2016 November, ??}
  \pagerange{\pageref{firstpage}--\pageref{lastpage}} \pubyear{2016}
 \maketitle
 \label{firstpage}

\begin{abstract}
We applied the currently most comprehensive version of the
statistical-parallax technique to derive kinematical parameters of
the maser sample with 136 sources. Our kinematic model comprises
the overall rotation of the Galactic disk and the spiral
density-wave effects. We take into account the variation of radial
velocity dispersion with Galactocentric distance. The best
description of the velocity field is provided by the model with
constant radial and vertical velocity dispersions, $(\sigma U0,
\sigma W0) \approx (9.4 \pm 0.9~, 5.9 \pm 0.8)~ km/s$. We compute
flat Galactic rotation curve over the Galactocentric distance
interval from 3 to 15 kpc and find the local circular rotation
velocity to be $ V_0 \approx (235-238)$~ km/s $\pm 7$~ km/s. We
also determine the parameters of the four-armed spiral pattern
(pitch angle $i \approx (-10.4 \pm 0.3)^\circ$ and the phase of
the Sun $\chi_0 \approx (125 \pm 10) ^\circ$). The radial and
tangential spiral perturbations are about $f_R \approx (-6.9 \pm
1.4)$~km/s, $f_\Theta \approx (+2.8 \pm 1.0$) ~km/s. The kinematic
data yield a solar Galactocentric distance of $R_0 \approx (8.24
\pm 0.12)~kpc$. Based on rotation curve parameters and the
asymmetric drift we Infer the exponential disk scale $H_D \approx
(2.7 \pm 0.2)$ ~kpc under assumption of marginal stability of the
intermediate-age disk, and finally we estimate the minimum local
surface disk density, $\Sigma (R_0) > (26 \pm 3) ~ M_\odot
pc^{-2}$.
\end{abstract}


 \section{INTRODUCTION}
The kinematics of Galactic disk populations has remained a focus
of persistent interest for many decades. The main task was and
remains the computation of the rotation curve of the disk, which
reflects the features of the mass distribution in the Galaxy and
is often used, especially by radio astronomers, to estimate the
kinematic distances to objects. To determine the run of the
rotation curve the researchers mostly used samples of young
Galactic-disk objects with bona fide distances (i.e., with
distances determined with relatively small random errors):
Cepheids, young open clusters, $HII$ clouds, star-forming regions,
OB stars, OB associations, red-clump giants, etc.
\citep{BurtonGordon1978,Fich1989,BrandBlitz1993,Damb1995,Glush1998,Glush1999,Melnik1999,
Rastor1999,Damb2001a,Damb2001b,Melnik2001a,Melnik2001b,Zabolot2002,
Bobylev2004,Bobylev2006,Bobylev2007,Bobylev2008,Melnik2009,Bobylev2009,Bobylev2013a}.
\cite{Demers2007} used carbon stars to trace the run of the
rotation curve beyond the solar circle out to a Galactocentric
distance of 15~kpc and then extended their study to the
Galactocentric distance of 24~kpc \citep{Battinelli2013}.
\cite{Bovy2012} performed a detailed analysis of the kinematics of
an extensive sample of red giants studied within the framework of
the SDSS-III/APOGEE project \citep{Eisenstein2011} in the
Galactocentric distance interval $4 < R < 14$~kpc. The authors of
a number of recent studies (e.g., \cite{Sofue2012} and
\cite{Bhat2014}) attempt to combine the disk rotation curve
constructed for the central region of the Galaxy with the
kinematic data for distant stars of the Galactic halo described by
stellar hydrodynamics equations.

\cite{Khop2003b} and \cite{Sofue2009} demonstrate the large
scatter of the circular velocities of different objects inferred by
different authors, which casts doubt on the reliability of
kinematic distances determined using the rotation curve. This
scatter can mostly be explained by the use of inconsistent
distance scales for different objects, different adopted values
of the solar Galactocentric distance, use of different techniques
for analyzing the space velocity field, and the neglect of the
effects associated with the deviation of the mass distribution
from axial symmetry (i.e., the influence of the bar and spiral
density waves).

Of great importance for modelling the Milky Way is the issue of
the effect of a massive Galactic bar on the kinematics of the
disk, primarily on the run of the rotation curve in the
Galactocentric distance interval $R < 4$~kpc. It is now perhaps
safe to say that the asymmetry of the $HI$ rotation curve first
found by \cite{Kerr1964, Kerr1969} using the tangent-point
method and confirmed by later studies, e.g., \cite{Levine2008}, is
due to this very effect. A simulation of the effect of the massive
Galactic bar on the results of the determination of the run of the
rotation curve in the central part of the Galaxy
\citep{Chemin2015} strongly supports this point of view. In
particular, the above authors showed that ignoring these effects
when decomposing the rotation curve may result in overestimating
the bulge mass by a factor of several times. In this case the
computed rotation curve in the central region within about 2~kpc
lies systematically above the ``true'' rotation curve.

The buildup of observational data for maser sources in Galactic
star-forming regions (see the review by \cite{Reid1981}) opened up
new prospects for a detailed analysis of the state and kinematics
of the thin Galactic disk. The authors of recent studies
\citep{Reid09, Stepan2011, Bajkova2012, Bajkova2013, Bobylev2013c,
Reid14, Bobylev2014a, Bobylev2014b, Bobylev2014c} used maser
sources to trace the rotation curve and estimate the parameters of
the spiral pattern of our Galaxy. In addition, \cite{Bobylev2013b}
estimated the pitch angle of the Milky-Way spiral pattern from the
space distribution of maser sources.

This study is also dedicated to the analysis of the kinematics of the currently largest sample
of Galactic masers, which is described in the Data section. Here we attempt to jointly determine
the rotation curve of the young disk and the systematic deviations from it caused
by the perturbations due to the spiral density wave. We use the method of statistical parallax
to refine the maser distance scale. To solve this problem, we propose a new method for
describing the field of space velocities, which allows the radial variation
of the velocity ellipsoid axes and the exponential scale of the disk to be estimated.

\section{DATA}
\label{Data}
We use published data for maser sources associated with very young stars located
in star-forming regions. Currently, high-precision astrometric VLBI measurements have been
performed by several research teams for more than one hundred such objects, yielding
trigonometric parallaxes and proper motions with errors that on the average are no greater than 10\%
and 1~mas/yr, respectively.

One of such observational campaigns is the Japanese
project VERA (VLBI Exploration of Radio Astrometry) targeting water ($H_2O$) Galactic masers
~\citep{Hirota07} and SiO masers -- which are very rarely found among young objects -- ~\citep{Kim08},
at 22 and 43~GHz, respectively.

Water and methanol (CH$_3$OH) masers are observed in the USA
(VLBA) at 22 GHz and 12 GHz, respectively \citep{Reid09}. Methanol
masers are also observed in the framework of the European VLBI
network~\citep{Rygl10}. These two projects are now parts of the
combined BeSSeL program \footnote
{http://www3.mpifr-bonn.mpg.de/staff/abrunthaler/BeSSeL/index.shtml}
(Bar and Spiral Structure Legacy Survey, \cite{BesseL11}).

Radio stars are observed with VLBI in the continuum at
8.4 GHz (Torres et al. 2007; Dzib et al. 2011) for the same purpose. In the framework of this program,
radio sources in the Local arm are observed that are mostly associated
with low-mass protostars \footnote {http://www.crya.unam.mx/$\sim$l.loinard/Gould/}.

The number of measured masers is rapidly increasing. The first
list of trigonometric-parallax, proper-motion, and radial-velocity
measurements contained the data for 18 sources \citep{Reid09}.

An analysis of this sample yielded a fairly large value of
the circular rotation speed, $V_0 \sim˜250$ ~km/s, and a
noticeable lag in circular rotation of star-forming regions ($\sim
15$ ~km/s) (Reid et al. 2009, Baba et al. 2009, Bovy et al. 2012,
McMillan \& Binney 2010). The influence of the spiral density
wave, especially visible in the radial velocities $V_R$ has been
found for a sample of 28 masers \citep{Bobylev2010}.

\cite{Honma2012} determined the fundamental parameters of the
Galaxy based on the data for 52 masers.

The most recent ``general'' review of astrometric measurements provides the data
for 103 masers \citep{Reid14}. It was followed by a number of publications by the
same authors dedicated to the analysis of masers located in individual spiral arms of
the Galaxy with refined parameters reported for some of the masers. \cite{Wu2014}
studied the Carina--Sagittarius arm (18 sources);
\cite{Choi14}, the Perseus arm (25 sources);
\cite{Sato2014}, the Scutum arm (16 sources);
\cite{Hachisuka2015}, the Outer arm (five sources),
and \cite{Sanna2014}, the inner region of the Galaxy (six sources).
Somewhat earlier,  \cite{Xu13} published their investigation of the Local arm (30 sources).

We supplemented the list of \cite{Reid14} with the results of 40
recent astrometric measurements, thereby increasing the sample of
masers with complete kinematic data to 136 sources. The references
are given in Table~5 of the Appendix.

 \section{THE METHOD}
 \label{The method}

\subsection{Basic ideas}
\cite{Wilson1991} were the first to use the method of statistical
parallax to study the kinematics of Galactic disk objects: they
applied it to analyze the velocity field and constrain the zero
point of the photometric distance scale of classical Cepheids
based on the PL relation. This was made possible by the
publication of the list of the then best Cepheid proper motions compiled by
\cite{KP}. However, the results of \cite{Wilson1991} were not very
accurate and in some aspects not too realistic (e.g., extremely
low, almost negligible vertical velocity dispersion of about
2~km/s) due to the small size of the sample (90 Cepheids) and
rather large proper-motion errors. The first post-Hipparcos
statistical-parallax study of Galactic-disk objects was performed
by \citet{Luri98}, who also used it to constrain the zero point of
the Cepheid distance scale. However, the above authors did not try
to analyze the kinematics of their Cepheid sample and used a
priori fixed values for Oort's constants and the Sun's
Galactocentric distance. The first bona fide post-Hipparcos
statistical-parallax study of Galactic disk objects combining both
the analysis of the velocity field and refinement of the
distance-scale zero point of classical Cepheids and young open
clusters was performed by \cite{Rastor1999}. It was followed by a
statistical-parallax analysis of OB-associations
\citep{Damb2001a}, early-type supergiants \citep{Damb2001b}, and a
mixed sample of various young luminous objects
\citep{Zabolot2002}.

In this study we describe the scatter of the difference between
the computed space velocity of an object, based on adopted
distance, and its model value by the matrix of covariances, which
is determined by random errors of observational data (heliocentric
distances, radial velocities, and proper motions), velocity
dispersion of the sample considered, and errors of model
velocities induced by random and systematic errors of heliocentric
distances. This approach, which allows the maximum-likelihood
method to be used to determine the kinematical parameters, was
first proposed and briefly described by~\cite{Murray83}. For
detailed step-by-step instructions see the electronic tutorial
by~\cite{Rastor02} and papers by~\cite{Hawley1986}
and~\cite{Damb2009}. We present the basic ideas of the method in
Sections~3.2 -- 3.5.

Given the large spatial extent of our sample we have to take into
account the variation of the parameters of the velocity ellipsoid
with Galactocentric distance. The ratio of the two axes of the
velocity ellipsoid parallel to the Galactic midplane obeys the
following Lindblad relation:

\begin{equation}
\label{Lindblad}
\frac{\sigma V (R)}{\sigma U (R)} =
\sqrt{1+\frac{R}{2\omega}\cdot\frac{d\omega}{dR}}.
\end{equation}

The thickness of the young disk varies little with Galactocentric
distance and therefore it is safe to assume, to a good first
approximation, that

\[
\sigma W (R) = \sigma W (R_0)= \sigma W0 = const.
\]
It only remains for us to decide how the radial velocity
dispersion of our maser sample may vary with Galactocentric
distance. We proceed from the results of numerical simulations of
the dynamical evolution of galaxy disks performed
by~\cite{Khop2003a}. The above authors showed that an initially
cool galactic disk rapidly reaches the state of marginal stability
where radial velocity dispersion reaches its limit value in
accordance with the criterion of~\cite{Toomre}. Note that
\cite{Sabur14} used the marginal stability assumption to estimate
the masses of galaxy disks.

The criterion of marginal stability written as
\[
\frac{\kappa (R)\cdot\sigma U (R)}{\Sigma (R)} > 3.36~G \
\]
should apply to the radial velocity dispersion of intermediate-age
stars, which are the main contributors to the surface density of
the disk. Needless to say it is greater than the velocity
dispersion of the youngest objects including Galactic masers.

As case (\textbf{A}), we further assume, that local velocity
dispersions of objects of different ages are proportional to each
other and to the surface density of the disk. Hence the condition

\[
\frac{\kappa (R)\cdot\sigma U (R)}{\Sigma (R)} \approx const~ (<
3.36 ~G), \
\]
where $\kappa (R)$ is the current epicyclic frequency; $\Sigma
(R)$, the surface density of the disk, and  $G$, the gravitational
constant, implies the following relation between the current
values of these parameters at Galactocentric distance $R$ and
the corresponding local values in the solar neighborhood (at the
Galactocentric distance $R_0$):

\begin{equation}
\label{Toomre}
\frac{\sigma U (R)}{\sigma U(R_0)} \approx
\frac{\kappa (R_0)} {\kappa (R)} \cdot \exp(\frac{R_0 - R}{H_D}),
\end{equation}
where $H_D$ is the exponential disk scale. Hence our proposed
first method of the description of the variation of radial
velocity dispersion with the Galactocentric distance makes it
possible to independently estimate the exponential disk scale
$H_D$ from kinematical data. Condition (\ref{Toomre}) can be
treated as some kind of ``equation of state'' for the sample
of objects under consideration.

In addition, we also examined two other cases of behavior of the
radial velocity dispersion: (\textbf{B}) proportionality of radial
velocity dispersion to the surface density of an exponential disk,

\begin{equation}
\label{SimpleExp}
 \frac{\sigma U (R)}{\sigma U(R_0)} \approx \exp(\frac{R_0 - R}{H_D}),
\end{equation}
and (\textbf{C}) constancy of radial velocity dispersion, i.e.
$\sigma U(R) = \sigma U (R_0) = \sigma U0 = const$.

\subsection{Kinematic models}
In this paper we consider two kinematical models. The first model includes
only differential rotation of the Galaxy with angular velocity $\omega (R)$ and solar motion
relative to the local sample:

\[
\left( {\matrix{
   {V_r}  \cr
   {k r \mu_l}  \cr
   {k r \mu_b}  \cr}} \right) -
   \left( {\matrix{
   {R_0 (\omega - \omega_0) \sin l \cos b}  \cr
   {(R_0 \cos l - r \cos b) (\omega - \omega_0) - r \omega_0 \cos b}  \cr
   {- R_0 (\omega - \omega_0) \sin l \sin b}
   }} \right) -
\]
\begin{equation}
\label{KinModel}
 - G^T \times
   \left( {\matrix{
   {U_0}  \cr
   {V_0}  \cr
   {W_0}  \cr}} \right) = \delta \vec V_{loc},
\end{equation}
where $V_r$ is the line-of-sight velocity; $k = 4.741$
km/s/kpc/(mas/yr); $r$, the adopted heliocentric distance of the
object (in kpc), $(\mu_l \; \mu_b)$, the components of proper
motion along Galactic coordinates (in mas/yr), and $(U_0 \; V_0 \;
W_0)$, the components of the velocity of the local sample relative
to the Sun, and $T$ denotes transposition.

The rotation matrix $G$

\begin{equation}
\label{Gmatrix}
   G = \left( \matrix{
   {\cos b \cos l} & {- \sin l} & {- \sin b \cos l}  \cr
   {\cos b \sin l} & {\cos l} & {- \sin b \sin l}  \cr
   {\sin b} & {0} & {\cos b}  \cr
} \right)
\end{equation}
transforms the components of the velocity of the object

\begin{equation}
\label{Vloc}
 \vec V_{loc} = \left( \matrix{ V_r \cr k r \; \mu_l \cr k r
\; \mu_b} \right),
\end{equation}
given in the local coordinate system (associated with the
direction to the object) into the velocity components

\[ \vec V_{gal} =
\left( \matrix{ U \cr V \cr W} \right) = G \times \vec V_{loc}
\]
in heliocentric Cartesian coordinate system $(x \; y \; z)$ with
the origin at the Sun, the $x$-axis pointing to the Galactic
center and the $z$-axis pointing to the North Galactic pole.

The second model also includes the  contribution of non-circular
motions induced by the spiral density wave and computed in linear
approximation:

\[
\left( {\matrix{
   {V_r}  \cr
   {k r \; \mu_l}  \cr
   {k r \; \mu_b}  \cr}} \right) -
   \left( {\matrix{
   {R_0 (\Omega - \Omega_0) \sin l \cos b}  \cr
   {(R_0 \cos l - r \cos b) (\Omega - \Omega_0) - r \Omega_0 \cos b}  \cr
   {- R_0 (\Omega - \Omega_0) \sin l \sin b}
   }} \right) -
\]
\[ - \left( {\matrix{
{-(R_0 (\frac{\Pi}{R} - \frac{\Pi_0}{R_0}) \cos l - \frac{\Pi}{R}
r \cos b) \cos b} \cr { R_0 (\frac{\Pi}{R} - \frac{\Pi_0}{R_0})
\sin l} \cr {(R_0 (\frac{\Pi}{R} - \frac{\Pi_0}{R_0}) \cos l -
\frac{\Pi}{R} r \cos b) \sin b} }} \right) -
\]
\begin{equation}
\label{FullKinModel}
 - G^T \times
   \left( {\matrix{
   {U_0}  \cr
   {V_0}  \cr
   {W_0}  \cr}} \right) = \delta \vec V_{loc},
\end{equation}
where $R$ and $R_0$ are the Galactocentric distances of the object
and the Sun, respectively; $\Pi$ and $\Pi_0$, the radial
perturbations of the velocity of the object and the Sun,
respectively. Here modified angular velocities
\begin{equation}
\label{ModOmega} \Omega = \omega + \frac{\Theta}{R}, \; \Omega_0
=\omega_0 + \frac{\Theta_0}{R_0},
\end{equation}
include tangential velocity perturbations of the object and the
Sun, respectively, and $\Pi, \Pi_0, \Theta, \Theta_0$ are given by
(\ref{P&Theta}).

In both models (\ref{KinModel},\ref{FullKinModel})   the
difference $\delta \vec V_{loc}$ between the observed and model
velocities is a random vector whose matrix of covariances is given
in Section~3.3.

\subsection{Matrix of covariances}
The maximum-likelihood method used in this study allows one to
determine not only the main kinematical properties of the sample,
but also infer the systematic correction to the underlying
distance scale of objects (\citealt{Murray83},
\citealt{Hawley1986}, \citealt{Rastor02}, \citealt{Damb2009}). We
further assume that the true distance $r_t$ is related to the
adopted distance $r$ as $r_t = r / P$, where $P$ is the
distance-scale correction factor.

The full matrix of covariances of vector $\vec V_{loc}$ includes errors of observational data;
``cosmic'' dispersion (three-dimensional distribution of residual velocities), and errors of
the model velocity field induced by random and systematic errors of the adopted distances \citep{Rastor02}:

\begin{equation}
\label{Lloc}
L_{loc} = L_{err} + L_{resid} + \delta L
\end{equation}

The observed local vector of velocity errors at the adopted
distance $r$ is given by the formula

\begin{equation}
\label{dVloc}
  \vec \delta V_{loc} = \left( \matrix{ \delta V_r \cr
  k r \; \delta \mu_l \cr
  k r \; \delta \mu_b } \right),
\end{equation}
and therefore in the absence of correlations between the errors of
line-of-sight velocities and those of proper-motion components
(which is the case for the sample of maser sources considered),
the matrix of observational errors has the form

\[ L_{err} = \langle \delta \vec V_{loc} \cdot \delta \vec
V_{loc}^T \rangle =
\]
\begin{equation}
\label{Lerr}
 = \left(
 \matrix{
   \sigma_{V_r}^2 & 0 & 0  \cr
   0 & k^2 r^2 \; \sigma_{\mu_l}^2 & k^2 r^2 \; \sigma_{\mu_l} \; \sigma_{\mu_b} \;
   \rho_{\mu_l \;
\mu_b} \cr
   0 & k^2 r^2 \; \sigma_{\mu_l} \; \sigma_{\mu_b} \; \rho_{\mu_l \;
\mu_b} & k^2 r^2 \; \sigma_{\mu_b}^2 \cr }
   \right),
\end{equation}
where angle brackets denote averaging over the ensemble of objects
in the given region and $\rho_{\mu_l \; \mu_b}$ is the correlation
coefficient of proper motion components.

\textbf{Note:} Data on maser sources are given in equatorial
coordinate system, with no correlations of proper motion
components. To simplify calculations, we first evaluate
the matrix of covariances $L_{err}$ in the equatorial coordinate system
and then use standard rotation transformation to transform the matrix of covariances to the galactic coordinate system.

Let us now introduce two auxiliary matrices:

\[
   P = \left( {\matrix{
   1 & 0 & 0  \cr
   0 & p & 0  \cr
   0 & 0 & p  \cr
 } } \right)
 \]
and

\[
   M = \left( {\matrix{
   0 & 0 & 0  \cr
   0 & 1 & 0  \cr
   0 & 0 & 1  \cr
 } } \right)
 \]

It is easy to understand that the first matrix relates the
components of space velocity $\vec V_{loc}$ computed for the
adopted heliocentric distances $r$ to the corresponding quantities
computed from the true heliocentric distances $r_t$ of objects.

We compute the matrix of covariances $L_{resid}$ that is
responsible for the intrinsic scatter of stellar velocities in the
Galaxy, which is usually described by the three-dimensional
Gaussian distribution with the principal axes $(\sigma U \; \sigma
V \; \sigma W)$, assuming, to a first approximation, that the
$(\sigma U \; \sigma V)$ axes are parallel to the Galactic
midplane with the major axis directed toward the Galactic center.
We take into account the inclination of the velocity ellipsoid in
the region of the Galaxy where we observe the object to the line
of sight. To this end we introduce the auxiliary angle $\varphi$
that determines the orientation of the velocity ellipsoid and
which is equal to the angle between the projection of the line of
sight onto the Galactic plane and the projection of the major axis
of the velocity ellipsoid. It can be easily shown that this angle
is given by the following formula

\[
\tan \; \varphi = {{R_0  \cdot \sin l} \over {R_0  \cdot \cos l -
r \cdot \cos b}}.
\]

The rotation matrix

\begin{equation}
\label{GsMatrix}
 G_S  = \left( \matrix{
   {\cos b\cos \varphi} & {\cos b\sin \varphi} & {\sin b}  \cr
   { - \sin \varphi} & {\cos \varphi} & 0  \cr
   { - \sin b\cos \varphi} & { - \sin b\sin \varphi} & {\cos b}  \cr
   } \right)
\end{equation}
determines the transformation from the coordinate system connected
with the axes of the velocity ellipsoid to the local coordinate
system connected with the direction to the object. In the
coordinate system determined by the principal axes of the velocity
ellipsoid the matrix of covariances has the following diagonal
form

\begin{equation}
\label{L0Tensor}
   L_0 = \left( {\matrix{
   \sigma U^2 & 0 & 0  \cr
   0 & \sigma V^2 & 0  \cr
   0 & 0 & \sigma W^2  \cr
 } } \right).
\end{equation}
When going to the local coordinate system it transforms in
accordance with the following well-known formula:

\begin{equation}
\label{Lturn}
L_{resid} = G_S \times L_0 \times G_S^T .
\end{equation}

Given the systematic correction to the adopted distance scale the
matrix of covariances $L_{resid}$ acquires the following final
form

\begin{equation}
\label{Lresid}
 L_{resid} = P \times G_S \times L_0 \times G_S^T \times P^T.
\end{equation}

The smallest contribution to covariance matrix is provided by the
third term, which is proportional to the squared random error of
the distance scale. This term includes the systematic velocity,
its derivative with respect to heliocentric distance, and the
matrix of ``cosmic'' dispersion:

\begin{equation}
\label{deltaL} \delta L = (\sigma_\pi / \pi)^2 \cdot [M \times \
G_S \times L_0 \times G_S^T \times M^T + \vec \Upsilon \cdot \vec
\Upsilon^T ],
\end{equation}
where to simplify the formulas we introduce the following auxiliary vector

\[
\label{Upsilon}
\vec \Upsilon = M \times [G^T \times \vec V_0 +
\vec V_{sys}] - r / p \cdot P \times \partial \vec V_{sys} /
\partial r.
\]
Here

\begin{equation}
\label{VSample}
\vec V_0 =
 \left( {\matrix{
   {U_0}  \cr
   {V_0}  \cr
   {W_0}  \cr}} \right)
\end{equation}
is the the velocity of the sample of objects relative to the Sun and

\begin{equation}
\label{Vsys}
\vec V_{sys} = \vec V_{rot} + \vec V_{spir}
\end{equation}
is the full velocity of systematic motions including the differential rotation
of the disk and perturbations from the spiral pattern. Note that
all components of the full matrix of covariances $L_{loc}$ are computed using
the adopted heliocentric distances to the objects.

\subsection{Systemic velocity field}

In our models we describe purely circular motions (the
differential rotation of the disk) via polynomial expansion of the
difference of angular velocities as a function of Galactocentric
distance in the following form

\begin{equation}
\label{DeltaOmega}
 (\omega  - \omega_0) \approx
 \sum \limits_{n=1}^M
 {\frac{1}{n!}{{\partial^n \omega_0} \over {\partial
 r^n}} \cdot (R-R_0)^n},
\end{equation}
with the order of expansion $M$ ranging from 4 to 5.

We describe the kinematical perturbations induced by the spiral
density wave in terms of the linear theory (\citealt{LinShu},
\citealt{LinYuanShu}). The trailing spiral pattern is described by
the following formula for the phase angle of the object relative
to the density wave:

\begin{equation}
\label{PhasChi}
 \chi - \chi_0 = m \cdot (\psi - \cot i \cdot \lg
\frac{R}{R_0}).
\end{equation}
Here $\chi_0$ is the phase of the Sun; $i$, the pitch angle (it is
negative for a trailing pattern); $\psi$, the Galactocentric
position angle of the object counted from the direction to the Sun
in the direction of the disk rotation, and $m$, the number of
spiral arms. Hereafter we set $m = 4$ given serious arguments for
the four-armed structure of the global spiral pattern provided by
a number of recent studies of the spiral pattern of the Galaxy
(\citealt{Vallee2013}, \citealt{Vallee2014}, \citealt{Vallee2015},
\citealt{Damb2015}).

The radial and tangential velocity perturbations can be written in
terms of the perturbation amplitudes $(f_R \; f_{\Theta})$ and
phase angles as follows:

\begin{equation}
\label{P&Theta}
 \left( {\matrix
 {{\Pi} \cr {\Theta}}} \right) =
 \left( {\matrix
 {{f_R \cdot \cos \chi} \cr {f_{\Theta} \cdot \sin \chi}}},
 \right), \;
\left( {\matrix
 {{\Pi_0} \cr {\Theta_0}}} \right) =
 \left( {\matrix
 {{f_R \cdot \cos \chi_0} \cr {f_{\Theta} \cdot \sin \chi_0}}},
 \right)
\end{equation}
where $\Pi_0$ and $\Theta_0$ are, respectively, the radial and
tangential perturbations for the Sun entering equations
(\ref{FullKinModel}, \ref{ModOmega}). Here $\Pi$ is positive in
the direction out of the Galactic center.

Note again that when computing the kinematical parameters of
the disk with the allowance for both the differential rotation and
perturbations induced by the spiral pattern the terms describing
differential rotation in the above formulas (the so-called
Bottlinger equations) should be modified as $\omega \rightarrow
\Omega = \omega + f_{\Theta} \cdot sin \chi / R, \omega_0
\rightarrow \Omega_0 = \omega_0 + f_{\Theta} \cdot sin \chi_0 /
R_0$ (see also \ref{ModOmega}), in order to explicitly single out
the tangential velocity component $f_{\Theta} \cdot sin \chi / R$
associated with perturbations.

The formula for the matrix of covariances includes the partial derivatives of the contribution
of systematic velocities with respect to heliocentric distance.
The corresponding formula for differential rotation can be easily derived:

\begin{equation}
\label{dVrotdr}
  {{\partial \vec V_{rot}} \over {\partial r}} = \left( {\matrix{
  {R_0 \cdot {\partial \over {\partial r}}(\omega - \omega _0 ) \cdot \sin l \cdot \cos b} \cr
  {(R_0 \cdot \cos l - r \cdot \cos b) \cdot {\partial \over {\partial r}}(\omega - \omega_0)-
  \omega \cdot \cos b} \cr
  { - R_0 \cdot {\partial \over {\partial r}}(\omega - \omega _0 ) \cdot \sin l \cdot \sin b} \cr
 }} \right) ,
\end{equation}
where the partial derivative of the difference of angular
velocities is equal to
\[
 {{\partial (\omega - \omega_0)} \over {\partial r}} \approx {\cos b \over R}
 \cdot (r \cos b - R_0 \cos l) \times
\]
\begin{equation}
\label{dOmegadr}
 \times \sum \limits_{n=1}^M
 {\frac{1}{(n-1)!}}{{\partial^n \omega_0} \over {\partial
 r^n}} \cdot (R-R_0)^{n-1},
\end{equation}
and $M$ is the order of expansion of the angular velocity into a
Taylor series.

The formula for the partial derivative of the contribution of the spiral pattern
to the velocity field is somewhat more complex. We first write simpler formulas for
$D$, derivatives of Galactocentric distance $R$, position angle $\psi$,
and phase angle $\chi$:

\[
  D = R_0 \cos l - r \cos b.
\]
\[
 {\partial {R} \over \partial {r}} = - D / R \cos b
\]
\[
 {\partial {\psi} \over \partial {r}} = R_0 / R^2 \cos b \sin l
\]
\begin{equation}
\label{dChidr}
 {\partial {\chi} \over \partial {r}} = m \cdot ({\partial {\psi}
 \over \partial {r}} - {\partial {R} \over \partial {r}} \cdot \cot i /
 R)
\end{equation}

The general formula for the column vector of the partial derivative

\begin{equation}
\label{dVspirdr}
  {\partial {\vec V_{spir}} \over \partial {r}} =
  \left( {\partial {V_r^{sp}} \over \partial {r}} \;
  {\partial {V_l^{sp}} \over \partial {r}} \;
  {\partial {V_b^{sp}} \over \partial {r}} \right)^T
\end{equation}
is too unwieldy to write and therefore we give the separate
formulas for its three components (line-of-sight velocities $V_r$,
tangential velocities along Galactic longitude and latitude $V_l,
\; V_b$).

\[
  {\partial {V_r^{sp}} \over \partial {r}} = f_R /R \cdot \cos b
  \cdot (\cos b \cos \chi +
\]
\[ + D \cdot
  (\cos \chi \cdot {\partial {R} \over \partial {r}} / R +
  \sin \chi \cdot {\partial {\chi} \over \partial {r}} )) -
\]
\[
  - f_\Theta R_0 / R \cdot \sin l \cos b \cdot
  (\cos \chi \cdot {\partial {\chi} \over \partial {r}} -
  \sin \chi \cdot {\partial {R} \over \partial {r}} / R ) ;
\]

\[
  {\partial {V_l^{sp}} \over \partial {r}} = - f_R  R_0 / R
  \cdot (\sin \chi \cdot {\partial {\chi} \over \partial {r}} +
  \cos \chi \cdot {\partial {R} \over \partial {r}} / R) \cdot \sin l +
\]
\[
  + f_\Theta /R \cdot ( \cos b \sin \chi + D \cdot
  ( \sin \chi \cdot {\partial {R} \over \partial {r}} / R -
  \cos \chi \cdot {\partial {\chi} \over \partial {r}} )) ;
\]

\[
  {\partial {V_b^{sp}} \over \partial {r}} =
  - {\partial {V_r^{sp}} \over \partial {r}} \cdot \tan b .
\]

\subsection{Distribution of residual velocities and the Likelihood Function}

The three-dimensional probability function of the distribution of
residual velocities  -- the difference between the observed and
model velocities of the star, $\delta \vec V_{loc}$ -- can be
written in the following general form \citet{Murray83}

\[
 f(\delta \vec V_{loc} \mid \Lambda) =
\]
\[ = (2\pi )^{-3/2}  \cdot \left|
 {L_{loc}} \right|^{ - 1/2}  \cdot \exp \{ -{1 \over 2} \cdot
 \delta \vec V_{loc}^T \times L_{loc}^{-1}\times \delta
 \vec V_{loc}\} ,
\]
where $\Lambda$ is the set of unknown parameters of the problem
that describe the model velocity field; $\left| {L_{loc} }
\right|$ and $L_{loc} ^{-1}$ are the determinant and inverse of
the matrix of covariances $L_{loc}$ (\ref{Lloc}), respectively,
computed individually for each object of the sample. Note that the
matrix of covariances also depends on the parameter vector
$\Lambda$. The distribution function has the meaning of the
probability density for the residual velocity of a particular
star. Because stars are distributed independently of each other in
the velocity space, their N-point distribution function is equal
to the product of functions $f_i(\delta \vec V_{loc} \mid
\Lambda)$ for all stars of the sample:

\begin{equation}
\label{dVlocDistrib} F(\delta \vec V_{loc}(1) , \cdots ,\delta
\vec V_{loc}(N) \mid \Lambda) = \prod\limits_{i = 1}^N {f_i(\delta
\vec V_{loc} \mid \Lambda)},
\end{equation}
where $N$ is the number of objects. The gist of the
maximum-likelihood method is that the real (i.e., the actual)
distribution of the velocities of the objects of our sample is
assumed to be the most probable among all possible distributions.
Hence the parameter vector $\Lambda$ (which in addition to the
parameters of the kinematical model includes the quantity $P$, the
distance-scale correction factor) responsible for the probability
density $f$ of residual velocities should be chosen so as to
\textit{maximize} the probability $F (\Lambda)$ for the actual
sample. This problem is usually solved by minimizing the logarithm
of the N-particle probability density $F$ with the sign reversed,
i.e., the so-called  \textit{likelihood function}

\[
 LF (\Lambda) =  - \ln F(\delta \vec V_{loc}(1), \cdots , \delta \vec
 V_{loc}(N) \mid \Lambda) =
\]
\[
   - \sum\limits_{i = 1}^N {\ln f(\delta \vec
 V_{loc}(i) \mid \Lambda)}
\]
and reducing the problem to the standard search for the
\textit{minimum of the likelihood function} $LF (\Lambda)$ using
one of the efficient \textit{multidimensional optimization}
algorithms. We now substitute into the above formula the formula
for probability density $f(\delta \vec V_{loc}(i) \mid \Lambda)$
to derive the following explicit formula for the likelihood
function:

\[
 LF (\Lambda) = {3 \over 2}N \cdot \ln 2\pi  +
\]
\begin{equation}
\label{LF}
 + {1 \over 2}\sum\limits_{i =
 1}^N [ \ln \left| {L_{loc}(i)} \right| +
 \delta \vec V_{loc}^T
 (i) \times L_{loc}(i) ^{ - 1} \times \delta \vec V_{loc}(i) ],
\end{equation}
where $i$-th covariation matrix $L_{loc}(i)$ is given by
(\ref{Lloc}), space velocity vector $\delta V_{loc}(i)$ is given
by (\ref{KinModel}) or (\ref{FullKinModel}) and summation is
performed over index $i$ denoting the current object of the
sample.

\subsection{Calculation of errors}
Probability function $LF (\Lambda)$ is a complicated nonlinear
function of the unknown kinematical parameters and the
distance-scale correction factor. In the vicinity of the global
minimum it can be approximated rather well by a multidimensional
quadratic function of all variables (many optimization methods are
based on such a representation of the target function). Strictly
speaking, the confidence intervals for the unknown kinematical
parameters (i.e., their root-mean-squared errors) can be
determined by projecting the section of the likelihood function
profile by the hypersurface $LF (\Lambda) = LF_0 +1$ (here $LF_0$
is the minimum value of the likelihood function reached in the
process of solution) onto the axes corresponding to parameters
$\Lambda$ \citep{NumRec}. With this, we can estimate not only the
errors of parameters determined but also their correlations (see
Figs. \ref{fig:R0_p_grey} and \ref{fig:R0_Om_grey} as examples).
The errors shown in the Tables 1 -- 4 were estimated by this
technique.

\begin{table*}
  \centering
  \caption{Solution for N = 131 maser sample (with Galactocentric distance $R \geq 3$ kpc).
  \textbf{Model A1}: circular rotation and spiral-wave perturbations (four arms) with 4th-order expansion
of the Galactic rotation curve. }\label{tab1}
\begin{tabular}{r r r r r r r r r r}
\hline
 $P$ & $R_0$ & $H_D$ & $U_0$ & $V_0$ & $W_0$ & $\sigma U0$ & $\sigma W0$\\
 & kpc & kpc  & km/s  & km/s & km/s & km/s & km/s  \\
\hline
0.979 & 8.21 & 4.31 & --11.06 & --18.26 & --8.76 & 10.02 & 5.74 \\
$\pm$ 0.018 & $\pm$ 0.12 & $\pm$ 0.90 & $\pm$ 1.31 & $\pm$ 1.19 & $\pm$ 1.06 & $\pm$ 0.90 & $\pm$ 0.75 \\
\hline \hline
$f_R$ & $f_\Theta$ & $\chi_0$ & $i$ & $\omega_0$ & $d \omega/dR$ & $d^2 \omega/dR^2$ & $d^3 \omega/dR^3$ & $d^4 \omega/dR^4$ & $LF_{min}$ \\
km/s & km/s & deg. & deg. & km/s/kpc & km/s/kpc$^2$  & km/s/kpc$^3$  & km/s/kpc$^4$ & km/s/kpc$^5$ \\
\hline
--6.80 & +3.10  & 123.1   & --10.44   & 28.94    & --3.91    & 0.86 & 0.01 & --0.08 & 1085.7870 \\
$\pm$ 1.37 & $\pm$ 0.95 & $\pm$ 10.1 & $\pm$ 0.29 &$\pm$ 0.51 & $\pm$ 0.07  & $\pm$ 0.03 & $\pm$ 0.03 & $\pm$ 0.10 & \\
\hline \hline
\end{tabular}
\end{table*}

\begin{table*}
  \centering
  \caption{\textbf{Model A2}: purely circular rotation and fixed scale coefficient $p = 1.00$ (5th-order expansion
of the Galactic rotation curve). }\label{tab2}
\begin{tabular}{r r r r r r r r r}
\hline
 $P$ & $R_0$ & $H_D$ &   $U_0$ & $V_0$ & $W_0$ & $\sigma U0$ & $\sigma W0$ \\
 & kpc & kpc  & km/s  & km/s & km/s & km/s & km/s  \\
\hline 1.00 & 8.31 & 4.34 & --7.73 & --17.69 & --8.64 & 11.59 & 5.65 \\
     & $\pm$ 0.13 & $\pm$ 0.75 & $\pm$ 1.52 & $\pm$ 1.20 & $\pm$ 0.91 & $\pm$ 0.95 & $\pm$ 0.80 \\
\hline \hline
$\omega_0$ & $d \omega/dR$ & $d^2 \omega/dR^2$ & $d^3 \omega/dR^3$ & $d^4 \omega/dR^4$ & $d^5 \omega/dR^5$ & $LF_{min}$ &  \\
km/s/kpc & km/s/kpc$^2$ & km/s/kpc$^3$ & km/s/kpc$^4$ & km/s/kpc$^5$ & km/s/kpc$^6$  &   &  \\
\hline
29.03 & --3.94  & 1.13  & --0.06  & --0.30  & 0.14  & 1465.8692 &  \\
$\pm$ 0.52 & $\pm$ 0.08 & $\pm$ 0.07 & $\pm$ 0.11 &$\pm$ 0.02 & $\pm$ 0.02  &  &  \\
\hline \hline
\end{tabular}
\end{table*}

\begin{table*}
  \centering
  \caption{\textbf{Model C1}: circular rotation and spiral-wave perturbations (four arms) with 4th-order expansion
of the Galactic rotation curve and constant velocity dispersions
$\sigma U$ and $\sigma W$. }\label{tab3}
\begin{tabular}{r r r r r r r r r r}
\hline
 $P$ & $R_0$ & $U_0$ & $V_0$ & $W_0$ & $\sigma U0$ & $\sigma W0$ \\
 & kpc & kpc  & km/s  & km/s & km/s & km/s \\
\hline
0.961 & 8.27 & --10.98 & --19.62 & --8.93 & 9.43 & 5.86 \\
$\pm$ 0.020 & $\pm$ 0.13 & $\pm$ 1.40 & $\pm$ 1.15 & $\pm$ 1.05 & $\pm$ 0.88 & $\pm$ 0.80 \\
\hline \hline
$f_R$ & $f_\Theta$ & $\chi_0$ & $i$ & $\omega_0$ & $d \omega/dR$ & $d^2 \omega/dR^2$ & $d^3 \omega/dR^3$ & $d^4 \omega/dR^4$ & $LF_{min}$ \\
km/s & km/s & deg. & deg. & km/s/kpc & km/s/kpc$^2$ & km/s/kpc$^3$  & km/s/kpc$^4$ & km/s/kpc$^5$ & \\
\hline
--7.00 & 2.62  & 130.3 & --10.39 & 28.35 & --3.83 & 1.17 & --0.08 & --0.30 & 1079.2939 \\
$\pm$ 1.48 & $\pm$ 1.05 & $\pm$ 10.8 & $\pm$ 0.25 &$\pm$ 0.45 & $\pm$ 0.08  & $\pm$ 0.05 & $\pm$ 0.04 & $\pm$ 0.03 & \\
\hline \hline
\end{tabular}
\end{table*}

\begin{table*}
  \centering
  \caption{\textbf{Model C2}: purely circular rotation with fixed scale coefficient ($p = 1.00$), 5th-order expansion
of the Galactic rotation curve and constant velocity dispersions
$\sigma U$ and $\sigma W$. }\label{tab4}
\begin{tabular}{r r r r r r r r r}
\hline
 $P$ & $R_0$ &  $U_0$ & $V_0$ & $W_0$ & $\sigma U0$ & $\sigma W0$ & \\
 & kpc & km/s  & km/s & km/s & km/s & km/s & \\
\hline
1.00 & 8.19 & --7.57 & --18.17 & --8.64 & 10.87 & 5.65 & \\
     & $\pm$ 0.12 & $\pm$ 1.55 & $\pm$ 1.20 & $\pm$ 0.92 & $\pm$ 0.91 & $\pm$ 0.80 & \\
\hline \hline
$\omega_0$ & $d \omega/dR$ & $d^2 \omega/dR^2$ & $d^3 \omega/dR^3$ & $d^4 \omega/dR^4$ & $d^5 \omega/dR^5$ & $LF_{min}$ &  \\
km/s/kpc & km/s/kpc$^2$ & km/s/kpc$^3$ & km/s/kpc$^4$ & km/s/kpc$^5$ & km/s/kpc$^6$ &  \\
\hline
28.64 & --4.00 & 1.28  & --0.10  & --0.37  & 0.19  & 1463.7012 &  \\
$\pm$ 0.53 & $\pm$ 0.09 & $\pm$ 0.04 & $\pm$ 0.02 &$\pm$ 0.02 & $\pm$ 0.02  &  &  \\
\hline \hline
\end{tabular}
\end{table*}

 \section{RESULTS}
 \label{Results}

\subsection{Kinematic parameters of the Milky Way young disk}

\begin{figure*}
\includegraphics[width=\linewidth]{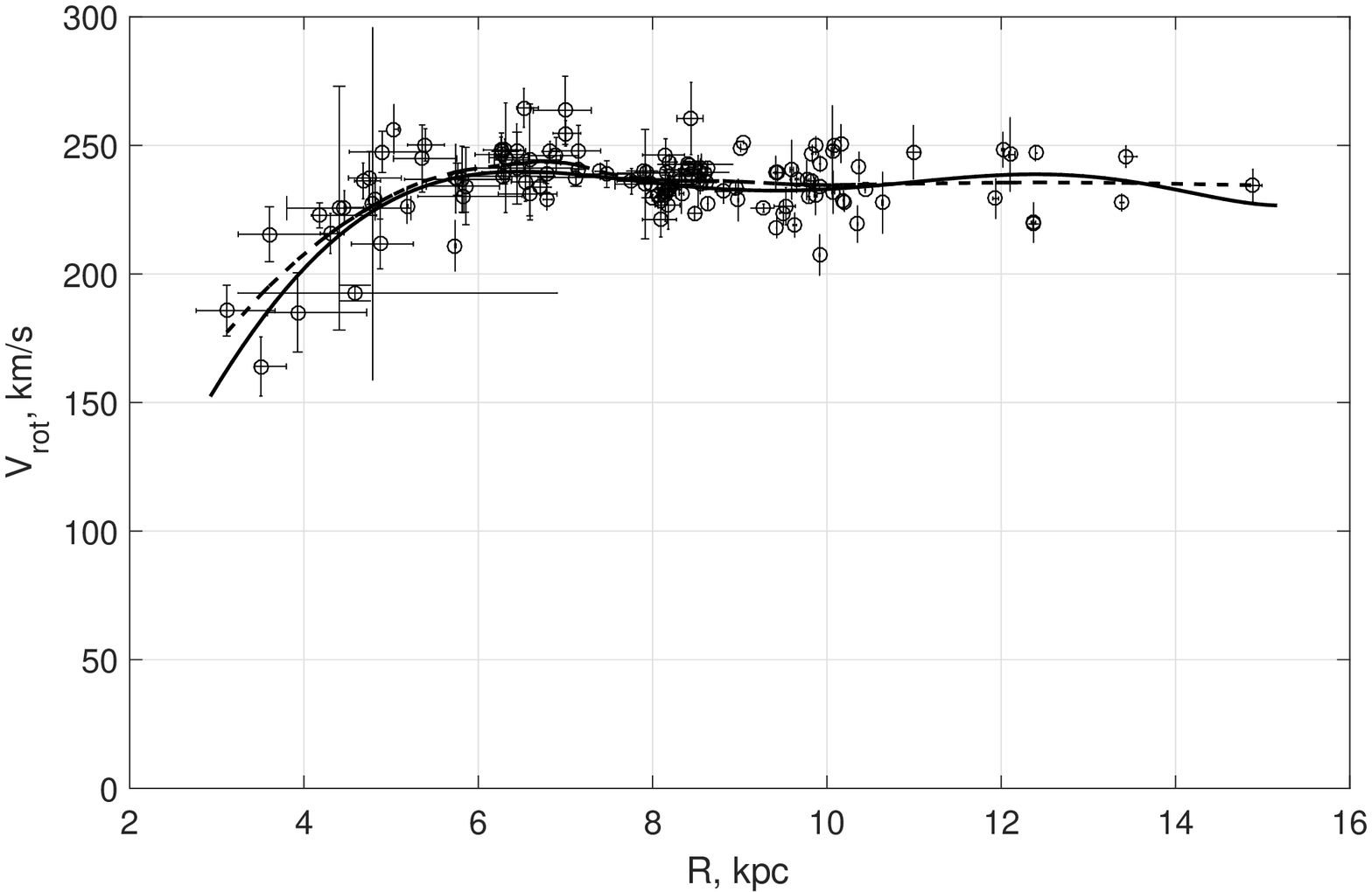}
\caption{Rotation curve of the Galactic disk. The solid line shows
the fit obtained by expanding angular velocity into a 4th-order
series in the distance difference (Model C1). Also plotted are the
tangential velocities of individual masers and their standard
errors computed based on the heliocentric distances, radial
velocities, and proper motions of the masers and the velocity of
the Sun inferred in this paper, as well as distance errors
computed based on parallaxes. The dashed line shows the smoothed
dependence of tangential velocities on Galactocentric distance
computed using the second-order locally weighted scatterplot
smoothing method (LOESS).}\label{fig:rotcurve}
\end{figure*}

We performed all our computations for masers with Galactocentric
distances greater than $3~$ kpc to reduce the effect of the
Galactic bar on the inferred kinematics of the sample
\citep{Chemin2015}. We further excluded from the initial list of
maser sources three objects whose (formally precise) observed
velocities differ from the corresponding model velocities by more
than $3~ \sigma$. The final sample consists of 131 maser sources.
The results are listed in the Tables 1 -- 4, where minimal values
of the likelihood functions, $LF_{min}$, are also shown for all
sets of calculations.

Tables~1 -- 4 list the inferred kinematical parameters and their
standard errors for four models: A1, A2, C1 and C2. Model A1
includes differential rotation and spiral perturbations
(\ref{FullKinModel}), whereas Model A2 includes only differential
rotation (\ref{KinModel}); both use Toomre-like ``equation of
state'' (\ref{Toomre}), i.e. dependence of radial velocity
dispersion on Galactocentric distance and disk surface density.
Our analysis of maser data in terms of these models yields an
exponential disk scale of be $H_D \approx (4.3 \pm
0.9)$ kpc.

Models B1 and B2 are similar to A1 and A2, respectively, but use
simple exponential dependence of radial velocity dispersion on
Galactocentric distance (\ref{SimpleExp}). These models imply
the disk exponential scales of $H_D \approx 18_{-10}^{+60}$ kpc
and $H_D \approx 16_{-6}^{+30}$ kpc, respectively. All other
parameters agree very well with those given by Models C1 and C2. As
we see from the analysis of errors, the likelihood function (LF)
profile along the $H_D$  axis near $LF_{min}$ is expected to be very
asymmetric, with a long ``tail'' extending to very large values of $H_D$. For
this reason, we do not present here these results.

We therefore decided to calculate a set of models C1 and C2,
similar to A1 and A2, respectively, but with $\sigma U(R) = \sigma
U0 = const$, i.e. with formally infinite values of the disc scale,
$H_D$, in (\ref{SimpleExp}). The corresponding model parameters are
listed in Tables 3 and 4. Systematically slightly lower values
of $LF_{min}$ for C1 and C2 with respect to A1 and A2 (by 6 and 2
units, respectively), indicate that Models C1 and C2 with constant
velocity dispersion along Galactocentric radius provide better fits
to observations with respect compared to the model with variable
$\sigma U$.

\begin{figure}
\includegraphics[width=\linewidth]{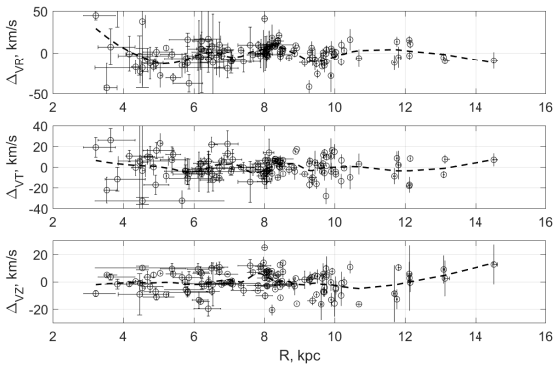}
\caption{Residual deviations of the components of maser space
velocities ($\Delta_{VR}, \Delta_{VT}, \Delta_{VZ}$) from the
Model C2 of purely circular rotation.}
\label{fig:Residuals_from_rotation}
\end{figure}

Figure~\ref{fig:rotcurve} shows the rotation curve of the
population of maser sources. It remains practically flat over the
interval from 5-6 to 15~kpc with a small depression at about 9 kpc
and small variations over $\sim$ 1 kpc scale lengths, which are
most likely due to spiral perturbations (see also
Fig.~\ref{fig:Residuals_from_rotation}). The computed rotation velocity at
Solar distance is of about $V_0 \approx (235 - 238)$~km/s
$\pm 7$~km/s.

\begin{figure}
\includegraphics[width=\linewidth]{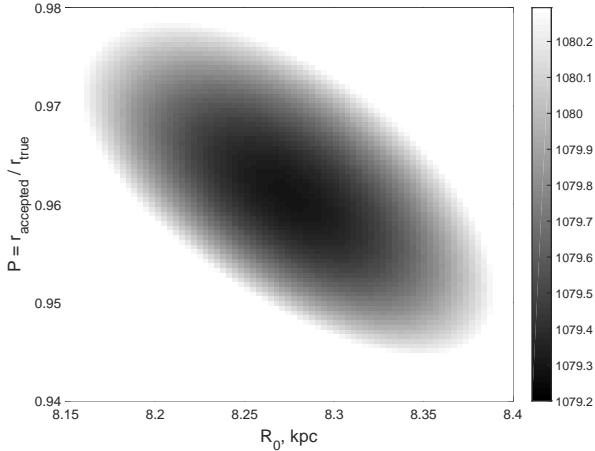}
\caption{Cross section of the likelihood function by the
hypersurface $LF = LF_0 + 1$ near the global minimum: projection
onto the $R_0$ and $P$ axes (Model C1).} \label{fig:R0_p_grey}
\end{figure}

\begin{figure}
\includegraphics[width=\linewidth]{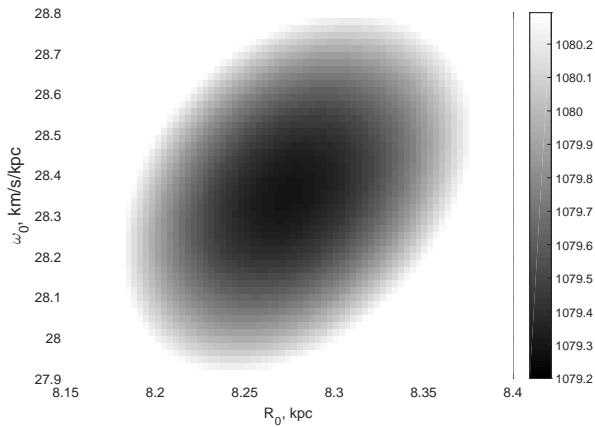}
\caption{Cross section of the likelihood function by the
hypersurface $LF = LF_0 + 1$ near the global minimum: projection
onto the $R_0$ and $\omega_0$ axes (Model C1).}
\label{fig:R0_Om_grey}
\end{figure}

Some of the parameters incorporated into our models of the
velocity field are mutually correlated.
Figures~\ref{fig:R0_p_grey} and~\ref{fig:R0_Om_grey} show the
($R_0, \; P$) and ($R_0, \; \omega_0$) two-parameter scattering
ellipses, respectively. It is easy to understand that $R_0$ also
correlates with the derivative of angular velocity, $(d\omega /
dR)_0$, and radial velocity dispersion, $\sigma U0$.

Figure~\ref{fig:Residuals_from_rotation} shows the residuals of
the radial, $\Delta_{VR}$, tangential, $\Delta_{VT}$, and
vertical, $\Delta_{VZ}$, velocity components from the model of
purely circular motions. Large quasi-periodic variations of radial
velocity, $\Delta_{VR}$, are immediately apparent, which are due to
perturbations produced by the spiral pattern. The $\Delta_{VT}$ residuals also
show similar but not so evident variations. Similar behaviour of
vertical component, $\Delta_{VZ}$, was first noted by
\cite{Bobylev2015}.

Models A1 and C1 provide very reliable estimates of the parameters
responsible for velocity perturbations induced by the spiral density wave.
Both the mean radial and tangential perturbation amplitudes are
significant and estimated as $f_R \approx (-6.9 \pm 1.4)$ km/s and
$f_{\Theta} \approx (+2.8 \pm 1.0)$ km/s, respectively. The mean phase angle of the
Sun relative to the ridge of spiral density wave is close to
$(125^\circ \pm 10)^\circ$, and the estimates of the pitch angle
are very stable and accurate: $i \approx (-10.4 \pm
0.3)^\circ$. The adequateness of the description of the spiral pattern is
additionally confirmed by very substantial reduction of $LF_{min}$
value (by approximately 380 -- 384 units) for Models A1 -- C1 compared to
Models A2 -- C2, respectively, which take into account
only pure rotation.

Our best Model C1 (Table 3) yields the estimates of radial and
vertical velocity dispersion for extremely young stellar
populations, $\sigma U0 \approx (9.4 \pm 0.9)$ km/s and $\sigma W0
\approx (5.9 \pm 0.8)$ km/s, which are less than the corresponding values for
young populations, such as open star clusters and Cepheids
\citep{Rastor1999,Damb2001b,Rastor2001,Zabolot2002}, and comparable
to the results obtained for very young OB-associations
\citep{Damb2001a,Melnik2009} and longest-periods Cepheids \citep{Bobylev2016}
based on GAIA Dr.1 data \citep{Gaiadr1}. Small
values of the velocity dispersions, $\sigma U$ and $\sigma W$,
indicate that maser sources, which are the representatives of the
``coolest'' disk population, retain the dynamical properties of
the interstellar medium. Consequently, the most likely ``equation of
state'' of maser population means that radial and vertical
velocity dispersions do not depend on Galactocentric distance.

Note again, that in our analysis of the kinematics of Galactic
masers we for the first time took into account the variation of
the shape of the velocity ellipsoid in accordance with the
Lindblad theorem.

\begin{figure}
\includegraphics[width=\linewidth]{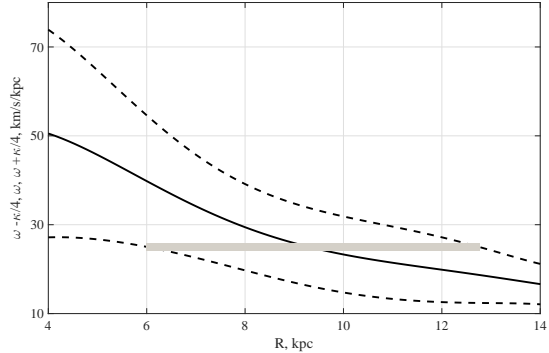}
\caption{Resonance diagram for the sample of Galactic masers computed for
the four-armed spiral pattern. The solid line shows the Galactocentric distance
dependence of the angular velocity of disk rotation  $\omega (R)$;
the dashed lines show the positions of the Lindblad resonances (the upper and lower curves
correspond to the outer ($\omega (R) + \kappa (R)/4$) and inner ($\omega (R) - \kappa (R) /4$)
resonances, respectively. The gray strip corresponds to the pattern
speed $\Omega_P$~$\approx$~25~km/s/kpc~\citep{Damb2015}.}
\label{fig:Resonances4_Op}
\end{figure}

We used our rotation curve and computed the resonance diagram for
four-armed spiral pattern (Fig.~\ref{fig:Resonances4_Op}).
This diagram demonstrates that a
global four-armed spiral pattern can exist in the Galactocentric
distance interval from 6 to 13~kpc with a corotation distance of
about 9.5 - 10 kpc and a pattern speed of about 25 km/s/kpc (see,
e.g., \cite{Damb2015}).

\begin{figure}
\includegraphics[width=\linewidth]{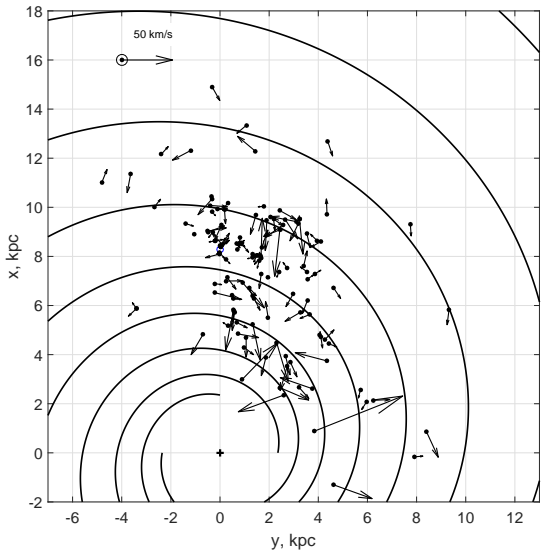}
\caption{Kinematic spiral pattern and positions of the masers of the sample
studied. The large pentagram shows the position of the Sun. The arrows show the deviations
of the space velocities of objects from purely circular orbits. The scale vector at the
top left corner corresponds to the velocity of  50 km/s.} \label{fig:Spirals}
\end{figure}

Figure~\ref{fig:Spirals} shows the positions of maser sources and the four
kinematic spiral arms determined in this study.
Note the concentration of masers to the Perseus, Carina-Sagittarius, and Inner arms,
and the small number of masers  near the Outer arm. The inferred parameters of the
spiral pattern -- including the pitch angle and the phase of the Sun --
agree well with recent results of \cite{Damb2015} obtained by analyzing
the space distribution of a large Cepheid sample.

\begin{figure}
\includegraphics[width=\linewidth]{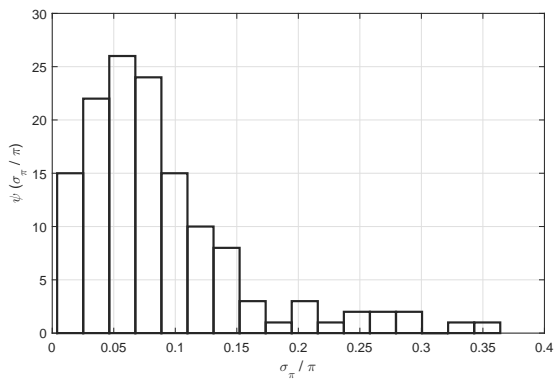}
\caption{The distribution of relative errors of measured masers
parallaxes.} \label{fig:RelPi}
\end{figure}

\begin{figure}
\includegraphics[width=\linewidth]{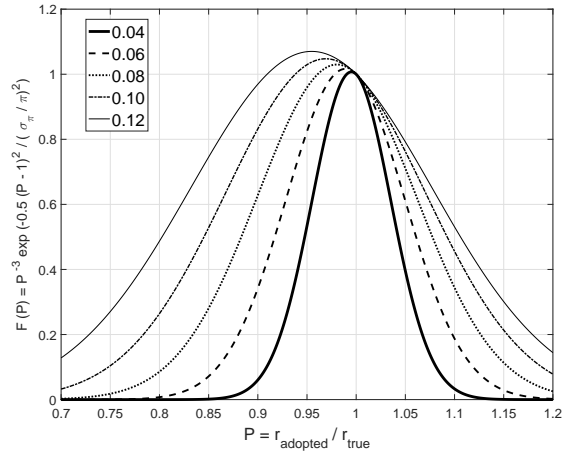}
\caption{Lutz-Kelker bias $F (P)$ (\ref{LK_equation}) calculated
for different values of the relative parallax error, $\sigma_{\pi}
/ \pi = 0.04, \; 0.06, \; 0.08, \; 0.10, \; 0.12$, for flat
stellar distribution. Our estimates of the distance scale
coefficient $P \approx 0.98 \pm 0.02$ and $0.96 \pm 0.02$ (Models
A1, C1) are fully consistent with the characteristic accuracy of maser
parallaxes (see also Fig. \ref{fig:RelPi}), with the mean and median
values of $\sigma_\pi / \pi \approx 0.09$ and $0.07$,
respectively.)}
 \label{fig:LK}
\end{figure}

\subsection{Maser distance scale}

First and foremost, we point out that the adjustment factor $P$
for the scale of maser distances is close to unity for Models A1
and C1. We calculated the Lutz-Kelker bias for 5 different levels of
relative errors of masers parallaxes, $\sigma_\pi / \pi = 0.04,
0.06, 0.08, 0.10, 0.12$, from modified standard expression
\citep{LK1973} that we adjusted to the case of flat distribution of
stars (see Fig.~\ref{fig:LK}):

\begin{equation}
\label{LK_equation} F (P) \sim P^{-3} exp [- \frac{(P-1)^2}{2
(\sigma_\pi / \pi)^2}],
\end{equation}
where $P = r / r_t = r_{adopted} / r_{true}$. The systematic shift of
the median and mean values of the distribution function $F (P)$ to the
left can be explained by systematic under-estimation of distances
calculated from parallax data. As can be seen, our estimates of
the distance scale factor $P \approx 0.98 \pm 0.02$ and $P \approx
0.96 \pm 0.02$ (Models A1 and C1 respectively) agree well with
the typical mean accuracy of maser parallaxes (see
Fig.~\ref{fig:RelPi}). Consequently, the scale of maser
trigonometric parallaxes has no large systematic biases despite
the fact that the mean fractional error of measured parallaxes is
of about 5-7\%. Considering this, we set $P = 1.00$ in
Models A2 and C2.

All models, A1 -- A2, C1 -- C2 yield very close estimates for the
Galactocentric distance of the Sun, $R_0$, with the weighted mean
value of $<R_0> \approx 8.24 \pm 0.12$ kpc, which agrees very well
with the most recent determinations.

\subsection{Disk scale estimate based on asymmetric drift law}

Let us now obtain an independent estimate for the disk scale $H_D$
for intermediate-age populations based on hydrodynamic equation of
stellar dynamics, observational data about the asymmetric drift
law in the local neighborhood \citep{DB1998} and inferred rotation-curve
parameters assuming that the
radial velocity dispersion, $\sigma V (R)$, follows Toomre-like
``equation of state'' (\ref{Toomre}).

We first write the well-known formula for the asymmetric drift in axisymmetric
stellar systems with small dispersion of residual velocities:

\begin{equation}
\label{AsymDrift} \Delta\Theta = \Theta_c - \Theta_0 \approx
\frac{\displaystyle \sigma U^2}{\displaystyle
2\Theta_0}\left(\frac{\displaystyle \sigma V^2}{\displaystyle
\sigma U^2} -R\frac{\displaystyle\partial}{\displaystyle\partial
R}\ln\left(R \cdot \nu \cdot\sigma U^2\right) \right),
\end{equation}
where $\Theta_0$ is the velocity of disk rotation at distance  $R$ from the rotation axis; $\Theta_c$,
the circular velocity related to the Galactic potential by the following formula
\[\Theta_c^2 = -R \frac{\partial\Phi (R,z)}{\partial R};
\]
$\sigma U^2$ and $\sigma V^2$, as above, are the radial and tangential dispersions of peculiar velocities, respectively, and $\nu$ is the disk volume density. Like in the above computations
we further assume that vertical velocity dispersion is constant and that the variation of the disk
thickness can be neglected. In this case the volume mass density is proportional to surface
density, $\nu (R) \sim \Sigma (R) \sim \exp (-R/H_D)$. Furthermore, when computing the derivative
\[
\frac{\displaystyle\partial}{\displaystyle\partial R}\ln\left(R
\cdot \nu \cdot \sigma U^2\right),
\]
we take into account the fact that variation of radial velocity dispersion $\sigma U^2$ with
Galactocentric distance is related to the variation of epicyclic frequency
\[
\kappa^2 = 4\omega(\omega + \frac{R\omega'}{2} ),
\]
and surface density  $\Sigma (R)$ of the disk by  Toomre-like
formula (\ref{Toomre}).

\cite{DB1998} and \cite{SBD2010} showed that asymmetric drift law
in the local neighborhood of the Galactic disk can be expressed as
\[
\Theta_c - \Theta_0 \approx \frac{\sigma U^2}{80 \; km/s} \equiv
\zeta \cdot \sigma U^2,
\]
where
\[
\zeta \approx \frac{\displaystyle 1}{\displaystyle 2\Theta_0}
\left( \frac {\sigma V^2}{\sigma U^2}
-R\frac{\displaystyle\partial}{\displaystyle\partial R}\ln\left(R
\cdot \nu \cdot \sigma U^2\right) \right) \approx
\frac{\displaystyle 1}{\displaystyle 80} \frac{\displaystyle
s}{\displaystyle km}.
\]
After simple manipulations with above expressions, we obtain
\begin{equation}
H_D \approx \frac{\displaystyle 3R}{\displaystyle 2\Theta_0 \;
\zeta + 1 - \frac{\sigma V^2}{\sigma U^2} - \frac{\displaystyle
5R\omega\omega'+ (\omega'^2+\omega\omega'')R^2}{\displaystyle
\omega \left(2 \omega + R \omega' \right)} }. \label{HD}
\end{equation}
We then substitute our inferred values of the Galactic rotation
curve parameters into expression (\ref{HD}) to estimate the radial
scale as

\begin{equation}
\label{HDBinney} H_D \approx (2.7 \pm 0.20) \; kpc;
\end{equation}
the standard error was calculated via Monte--Carlo method taking
into account all errors of other parameters that appear in the
formula for the disk scale $H_D$.

This value of the exponential scale of intermediate-age disk agrees
well with the \cite{DB1998} estimate based on simple exponential
variation of radial velocity dispersion; with the dynamic estimates
obtained by \cite{Khop2003b} ($\sim 3~$ kpc), \cite{McMillan2011}
($3.0 \pm 0.22$ kpc), \cite{Reid14} ($2.44 \pm 0.16~ kpc$),
\cite{McGaugh2016} ($2.0 - 2.9~kpc$), and with the results of an
analysis of the space distribution of stars based on SDSS data
performed by \cite{Juric2008} ($2.6 \pm 0.52$ kpc). Note, however,
that our estimates are appreciably smaller than that
obtained by \cite{Benjamin2005} ($3.9 \pm 0.6$ kpc) based on the
data about the distribution of stars of the GLIMPSE IR survey, and
somewhat greater than the recent model dynamic estimate obtained
by \cite{Bovy2013} ($2.15 \pm 0.14$ kpc), which is based, among
other things, on the data about the space distribution of stars.

Our estimates of the exponential scale of maser disk based on the
assumption of a Toomre-like ``equation of state'' (\ref{Toomre}),

\begin{equation}
\label{HDours} H_D \approx (4.3 \pm 0.8) \; kpc;
\end{equation}
(see Tables 1 and 2) are appreciably larger than (\ref{HDBinney}).
The maser disk seems to be more homogeneous along Galacticentric
radius, and Models C1 -- C2 provide a better description of the young
disk kinematics. It is appropriate to note that estimate
(\ref{HDBinney}) involves local properties of the velocity field,
whereas (\ref{HDours}) uses only global kinematics of maser
sample.

\subsection{Local surface density of the thin disk}

We can now use the assumption that the intermediate-age Galactic
disk is marginally stable to estimate the minimum local surface
disk density by the Toomre criterion (evidently, taking into
account the fact that the average radial velocity dispersion of
most of the disk stars is appreciably higher than the velocity dispersion
of masers ($\sim 10$~ km/s), which represent the youngest
population of the Galactic disk):

\[
\Sigma (R_0) > \frac{\kappa (R_0) \cdot \sigma U_0}{3.36 G}
\approx (26 \pm 3)~ M_\odot pc^{-2}.
\]

The standard error is estimated using Monte--Carlo technique based
on our inferred errors of other parameters. The characteristic
radial velocity dispersion for the sample of classical Cepheids
younger than  150 Myr is $(13 - 15)$~ km/s
\citep{Zabolot2002,Bobylev2016}, and it is even higher for older
stars. It is therefore safe to assume that the total local surface
disk density  should be at least twice greater than the above
estimate and amount to $(50 - 60)~ M_\odot pc^{-2}$.

Local surface density was estimated in many studies. Let us
mention only the relatively recent ones. Thus \cite{Korch2003}
reported a broad interval of local surface density estimates based
on the distribution of old red giants: $(10 - 42) \pm 6)~ M_\odot
pc^{-2}$, with the local volume mass density equal to  $\sim 0.1~
M_\odot pc^{-3}$. \cite{Bien2006} estimate the local surface
density to be $(57 - 79)~ M_\odot pc^{-2}$ within the $1.1~
kpc$-thick layer. \cite{Flynn2006} estimate the local surface density
as $\sim 49~ M_\odot
pc^{-2}$ as a result of their studies involving modelling of the
mass-to-luminosity ratio of our Galaxy $\sim 49~ M_\odot pc^{-2}$.
\cite{McGaugh2016} obtained a dynamic estimate of $(34 -
61)~M_\odot pc^{-2}$ for the average surface density at the solar
ring based on their modeling of the  terminal velocity curve.
Finally, \cite{Joshi} estimate the local surface mass density as
$\sim 40~\pm~12~ M_\odot pc^{-2}$
from their estimate of the scaleheight of open clusters younger
than 0.8~GYr and located within 0.4--2.0~kpc heliocentric distance
interval and combined with adopted Oort's constant estimates of $A$~=~14.8~km/s/kpc
and $B$~=~14.8~km/s/kpc. It
can be easily seen that our lower estimate of local surface
density agrees well with other determinations.

\section{CONCLUSIONS}

We analyzed the kinematics of the youngest disk population using
the currently most extensive sample of Galactic masers. This is the
first time that the method of statistical parallaxes in its most
general form (\citep{Murray83,Rastor02,Damb2009}) is used to study
the kinematics of the youngest population of the Galactic disk. We
applied this method to a sample of 131 maser sources located in
star-forming regions. The proposed method most adequately accounts
for all errors in the initial observational data (random and
systematic distance errors, random errors of line-of-sight
velocities and proper motions), as well as systematic (rotation of
the disk and perturbations due to the density wave) and random
(ellipsoidal distribution of residual space velocities) motions in
the sample, and the errors of the model velocity field due to the
random and systematic errors of object distances. As a result, the
distribution of the difference between the observed and model
velocities of each object is described by the matrix of
covariances, which includes both the observed quantities and the
full set of parameters to be determined. To determine the unknown
parameters we used the method of minimization of the
likelihood function.

Because of the large radial extent of the sample we had to take
into account the variation of the form and size of the ellipsoid
of residual velocities with Galactocentric distance. To this end,
we considered three cases: \textbf{(A)} Toomre-like ``equation of
state'' (\ref{Toomre}), \textbf{(B)} simple exponential decrease
of radial velocity dispersion (\ref{SimpleExp}) and \textbf{(C)}
constant radial velocity dispersion. We further assume that the
ratio of the two horizontal axes of the ellipsoid of residual
velocities obeys the Lindblad relation (\ref{Lindblad}), i.e., is
determined by the current local values of angular velocity and
epicyclic frequency.

The main results of this study are:

(1) The distance scale of maser sources based on VLBI
trigonometric parallaxes requires practically needs no systematic
correction despite rather substantial random errors of maser
parallaxes amounting in the average to $5 - 7\%$.

(2) The maximum-likelihood method yields the following estimates of
the main parameters of the Galactic disk: solar Galactocentric
distance ($8.24 \pm 0.12$)~ kpc, mean components of the maser
sample relative the Sun, ($U_0 \approx -11.0 \pm 1.3, \; V_0
\approx -19.0 \pm 1.2, \; W_0 \approx -9.0 \pm 1.1$)~km/s. The best
fit to the data is provided by Model C with constant velocity
dispersions, ($\sigma U0 \approx 9.4 \pm 0.9, \; \sigma W0 \approx
5.9 \pm 0.8$)~km/s.

We determined the rotation curve of the Galactic disk over the
Galactocentric distance interval $3 - 15$~kpc and found the
rotation velocity at the solar distance to be $(235 - 238) \pm 7$
~km/s, which agrees well with the results of
\cite{Reid14}. The rotation curve remains practically flat from 5
-- 6 out to 15 ~kpc.

(3) We determined the pitch angle $(-10.4~ \pm 0.3)^\circ$ and the
phase of the Sun $(125 \pm 10)^\circ$ of the four-armed trailing
spiral pattern in terms of the linear density-wave theory
(\citep{LinShu, LinYuanShu}). These estimates are in excellent
agreement with the results of the investigation of the space
distribution of classical Cepheids in the Galaxy \citep{Damb2015}.
We calculate the radial and tangential amplitudes of the perturbations
due to the spiral pattern ($f_R \approx -6.9 \pm 1.2, \; f_\Theta
\approx +2.8 \pm 1.0$) ~km/s. According to our data, the global
spiral pattern with a pattern speed of $\Omega_P \sim 25$~
km/s/kpc \citep{Damb2015} may exist in the Galactocentric distance
interval $6 - 13~ kpc$ with the corotation near $9.5 - 10$ ~kpc.

(4) We used Jeans hydrodynamic equations and detailed data about
the kinematics of stars in the local neighborhood
\citep{DB1998,SBD2010}, i.e., about the ``lag'' of centroids of
flat subsystems relative to the LSR, to independently estimate the
exponential scale of intermediate-age disk. The result $H_D
\approx (2.7 \pm 0.20)$ ~kpc agrees well with others estimates
made by different methods. Based on these considerations we
obtained a lower estimate for the disk surface density, $(26 \pm
3)~ M_\odot pc^{-2}$, which, on the whole, is consistent with
other published estimates of the total local surface density.

\section*{ACKNOWLEDGEMENTS}

A.T.~Bajkova and V.V.~Bobylev acknowledge the support from the
Presidium of the Russian Academy of Sciences (Program P--41
``Transitional and explosive processes in astrophysics'').
A.S~Rastorguev, N.D.~Utkin, M.V.~Zabolotskikh, and A.K.~Dambis
acknowledge the support of the analysis of maser kinematics from
the Russian Science Foundation (grant no.~14-22-00041) and the
support of the acquisition of observational data from the Russian
Foundation for Basic Research (grant no.~14-02-00472). A.K.~Dambis
acknowledges the support from joint grant by the Russian
Foundation for Basic Research and Department of Science and
Technology of India through project no. RFBR 15-52-45121 and
INT/RUS/RFBR/P-219. The authors are very grateful to the anonymous
referee for extremely valuable notes and criticism which made this
paper more readable and well-grounded.

\newpage
\section*{Appendix}.

Table~5 provides the data for 38 additional maser sources not
included in the last list of \cite{Reid14}. The table contains the
J2000 equatorial coordinates, trigonometric parallaxes $\pi$ and
their errors $\sigma_\pi$, proper motion components $\mu_\alpha,\;
\mu_\delta$ and their errors $\sigma_\alpha, \; \sigma_\delta$,
radial velocities $V_{LSR}$ relative to the LSR and their errors
$\sigma_{Vr}$. The last column gives the references to individual
maser data.

Note: The last two maser sources in the Table~5 were presented in
original list of \cite{Reid14}, but here we present improved data
adopted from new studies.

References in Table 5:\\
\\
1 - \cite{Choi14};\\
2 - \cite{Imai2012};\\
3 - \cite{Chib2014};\\
4 - \cite{Kusuno2013};\\
5 - \cite{Sakai2014};\\
6 - \cite{Xu13};\\
7 - \cite{Dzib2010};\\
8 - \cite{Honma2012};\\
9 - \cite{Burns2014};\\
10 - \cite{Reid2011};\\
11 - \cite{M-J2009};\\
12 - \cite{Burns2014a};\\
13 - \cite{Dzib11};\\
14 - \cite{Torres2012};\\
15 - \cite{Torres2007};\\
16 - \cite{Torres2009};\\
17 - \cite{Loinard2007};\\
18 - \cite{Kamez2014};\\
19 - \cite{Kim08};\\
20 - \cite{Reid14a};\\
21 - \cite{Naga2015};\\
22 - \cite{Motogi2015};\\
23 - \cite{Krish2015};\\
24 - \cite{Nakan2015};\\
25 - \cite{Burns2015};\\
26 - \cite{Dzib2016};\\
27 - \cite{Xu2016};\\
28 - \cite{Nakan2015};\\
29 - \cite{Burns2016};\\
30 - \cite{OrLe2016};\\
31 - \cite{Krish2016}.

\FloatBarrier
\begin{table*}[]
\caption{Additional list of 40 maser sources.}
\begin{tabular}{l l l l l r r r r r r r }
  \hline
  Name & RA J2000 & Decl J2000 & $\pi$ & $\sigma_\pi$ & $\mu_\alpha$ & $\sigma_\alpha$ &  $\mu_\delta$ & $\sigma_\delta$ &$V_{LSR}$& $\sigma_{Vr}$ & Ref.$\sharp$ \\
  & hh mm ss & dd mm ss & \multicolumn{2}{c}{mas} & \multicolumn{2}{c}{mas/y} & \multicolumn{2}{c}{mas/y} & \multicolumn{2}{c}{km/s} \\
  \hline
  G170.66--00.25 IRAS 05168+3634 & 05 20 22.07 & +36 37 56.6 & 0.537 & 0.038 &   0.23 & 1.07 &  --3.14 & 0.28 & --15.5 & 1.9 &   1  \\
  G108.43+00.89 IRAS 22480+6002 & 22 49 58.87 & +60 17 56.7 & 0.400 & 0.025 &  --2.58 & 0.33 &  --1.91 & 0.17 & --50.8 & 3.5 &   2  \\
  G110.19+02.47 IRAS 22555+6213 & 22 57 29.81 & +62 29 46.9 & 0.314 & 0.070 &  --2.04 & 0.35 &  --0.66 & 0.36 & --63.0 & 6.0 &   3  \\
  G115.06--00.05 PZ Cas          & 23 44 03.28 & +61 47 22.2 & 0.356 & 0.026 &  --3.70  & 0.20  &  --2.00  & 0.30  & --36.2 &
  0.7  &  4 \\
  G119.80--06.03 IRAS 00259+5625 & 00 28 43.51 & +56 41 56.9 & 0.412 & 0.123 &  --2.48 & 0.32 &  --2.85 & 0.65 & --38.3 & 3.1 &  5  \\
  G031.56+05.33 EC 95 Serp      & 18 29 57.89 & +01 12 46.1 & 2.291 & 0.038  &   3.599 & 0.026 &  --8.336 & 0.030 &   9.0  & 3.0  &  6, 7, 8, 30  \\
  G071.31+00.83 IRAS 20056+3350 & 20 07 31.25 & +33 59 41.5 & 0.213 & 0.026 &  --2.62 & 0.33 &  --5.65 & 0.52 &   9.0 & 1.0 &   9  \\
  G071.33+03.07 Cyg X-1         & 19 58 21.67 & +35 12 05.7 & 0.539 & 0.033 &  --3.78 & 0.06 &  --6.40 & 0.12 &  13.1 & 5.0  &  10 \\
  G073.12--02.09 V404 Cyg        & 20 24 03.82 & +33 52 01.9 & 0.418 & 0.024 &  --5.04 & 0.02 &  --7.64 & 0.03 &  16.9 & 2.2 &  11 \\
  G074.56+00.85 IRAS 20143+3634 & 20 16 13.36 & +36 43 33.9 & 0.367 & 0.037 &  --2.99 & 0.16 &  --4.37 & 0.43 &  --1.0  & 1.0  &  12 \\
  G109.87+02.11 Cep A  HW9      & 22 56 18.64 & +62 01 47.8 & 1.208  & 0.05  &  --1.03 & 0.10 &  --2.62 & 0.27 & --10.0 & 3.0 &  13, 27 \\
  G158.06--21.42 L 1448 C        & 03 25 38.88 & +30 44 05.2 & 4.31  & 0.33  &  21.90  & 0.07 & --23.10  & 0.33 &   4.5 & 3.0  &  6  \\
  G158.35--20.56 SVS13/NGC 1333  & 03 29 03.72 & +31 16 03.8 & 4.25  & 0.32  &  14.25 & 1.00  &  --8.95 & 1.40  &   7.5 & 5.0  &  6  \\
  G168.22--16.34 V773 Tau        & 04 14 12.92 & +28 12 12.3 & 7.70  & 0.19  &   8.30  & 0.50  & --23.60  & 0.50  &   7.5 & 0.5 &  14 \\
  G168.84--15.52 Hubble 4        & 04 18 47.03 & +28 20 07.4 & 7.53  & 0.03  &   4.30 & 0.05 & --28.90 & 0.30  &   6.1 & 1.7 &  15  \\
  G169.37--15.03 HDE 283572      & 04 21 58.85 & +28 18 06.4 & 7.78  & 0.04  &   8.88 & 0.06 & --26.60  & 0.10  &   6.0 & 1.5 &  15  \\
  G175.73--16.24 HP Tau/G2       & 04 35 54.16 & +22 54 13.5 & 6.20  & 0.03  &  13.85 & 0.03 & --15.40  & 0.20  &   6.8 & 1.8 &  16  \\
  G176.23--20.89 T Tau N         & 04 21 59.43 & +19 32 06.4 & 6.82  & 0.03  &  12.35 & 0.04 & --12.80 & 0.05 &   7.7 & 1.2 &  17  \\
  G203.32+02.05 NGC 2264        & 06 41 09.86 & +09 29 14.7 & 1.356 & 0.098 &  --1.00  & 0.60  &  --6.00  & 3.00  &   7.1 & 3.0 & 18 \\
  G208.99--19.38 Orion KL        & 05 35 14.51 & --05 22 30.5 & 2.39  & 0.030 &   9.56 & 0.10 &  --3.83 & 0.15 &   5.0 & 5.0 &  19  \\
  G353.02+16.98 DoAr21          & 16 26 03.02 & --24 23 36.4 & 8.20  & 0.37  & --26.47 & 0.92 & --28.23 & 0.73 &   3.0  & 3.0  &  6  \\
  G353.10+16.89 S1              & 16 26 34.17 & --24 23 28.5 & 8.55  & 0.50  &  --3.88 & 0.69 & --31.55 & 0.50 &   3.0  & 3.0  &  6  \\
  G353.94+15.84 IRAS 16293-2422 & 16 32 22.85 & --24 28 36.4 & 5.6   & 1.1   & --20.60  & 0.70  & --32.40  & 2.00   &  4.4 & 5.0  &  6  \\
  G045.37--00.22 GRS 1915+105    & 19 15 11.54 & +10 56 44.7 & 0.116 & 0.024 &  --3.19 & 0.03 & --6.24  & 0.05 &  30.4 & 1.0 &  20 \\
  G048.99--00.30 AGAL048.99-0.29 & 19 22 26.13 & +14 16 39.1 & 0.178 & 0.017 &  --2.16 & 0.09 & --5.87  & 0.17 &  66.3 & 0.3 &  21 \\
  G353.27+00.64 NGC 6357        & 17 26 01.59 & --34 15 14.9 & 0.59  & 0.06  &   0.47 & 0.07 &  0.99  & 1.04 &  --5.0 & 5.0 &  22  \\
  G339.88--01.26                 & 16 52 04.67 & --46 08 34.4 & 0.48  & 0.08  &  --1.60  & 0.10  & --1.90   & 0.10  & --38.8 & 5.0 & 23 \\
  G095.30--00.94 IRAS 21379+5106 & 21 39 40.80 & +51 20 35.0 & 0.262 & 0.031 &  --2.74 & 0.08 & --2.87  & 0.18 & --42.3 & 0.2 & 24, 28 \\
  G173.72--02.70 S235AB-MIR     & 05 40 53.38 & +35 41 48.4 & 0.63 & 0.03 &  0.79 & 0.12 & --2.41  & 0.14 & --17.91 & 3.1 & 25 \\
  G213.88--11.84 Mon2           & 06 10 50.59 & -06 11 50.4 & 1.12 & 0.05 &  --5.32 & 0.07 & 0.50  & 0.10 & 11.0 & 1.0 & 26 \\
  G213.70--12.60 Mon R2         & 06 07 47.86 & -06 22 56.5 & 1.166 & 0.021 &  --1.25 & 0.09 & 2.44  & 0.28 & 10.0 & 3.0 & 27 \\
  G054.10--00.08                & 19 31 48.80 & +18 42 57.1 & 0.231 & 0.031 &  --3.13 & 0.48 & --5.57  & 0.48 & 40.0 & 3.0 & 27 \\
  G058.77+00.64                & 19 38 49.13 & +23 08 40.2 & 0.299 & 0.040 &  --2.70 & 0.10 & --6.10  & 0.21 & 33.0 & 3.0 & 27 \\
  G059.47--00.18                & 19 43 28.35 & +23 20 42.5 & 0.535 & 0.024 &  --1.83 & 1.12 & --6.60  & 1.12 & 26.0 & 3.0 & 27 \\
  G059.83+00.67                & 19 40 59.29 & +24 04 44.2 & 0.253 & 0.024 &  --2.92 & 0.07 & --6.03  & 0.05 & 34.0 & 3.0 & 27 \\
  G071.52--00.38               & 20 12 57.89 & +33 30 27.1 & 0.277 & 0.013 &  --2.48 & 0.04 & --4.97  & 0.07 & 11.0 & 3.0 & 27 \\
  G192.60--00.04 S255IR-SMA1   & 06 12 54.02 & +17 59 23.3 & 0.563 & 0.036 &  --0.13 & 0.20 & --0.06  & 0.27 & 6.0 & 5.0 & 29 \\
  G108.18+05.51 L 1206       & 22 28 51.41 & +64 13 41.2 & 1.101 & 0.033 &  0.16 & 0.09 & --2.17  & 0.35 & --11 & 3.0 & 27 \\
  G305.200+0.019             & 13 11 16.93 & --62 45 55.1 & 0.25  & 0.06  &  --6.69 & 0.03 & --0.60  & 0.14 & --33.1 & 3.0 & 31 \\
  G305.202+0.208             & 13 11 10.49 & --62 34 38.8 & 0.25  & 0.05  &  --7.14 & 0.17 & --0.44  & 0.21 & --44.0 & 3.0 & 31 \\

  \hline
\end{tabular}
\end{table*}
\FloatBarrier


\end{document}